\documentclass
[twocolumn,aps,prd,amsmath,showpacs,floatfix,nofootinbib]
{revtex4-1}

\usepackage{CJK}                     
\usepackage[dvips]{graphicx}
\usepackage{bm}                      
\usepackage{mathptmx}                
\usepackage{dcolumn}                 
\usepackage{xcolor}                  %

\def\bea{\begin{eqnarray}}
\def\eea{\end{eqnarray}}
\def\be{\begin{equation}}
\def\ee{\end{equation}}
\def\non{\nonumber}

\def\anss{ans\"atze} 

\def\al{\alpha}
\def\eps{\varepsilon}

\def\rob{\rho_B}

\begin{document}

\title{
Hybrid neutron stars with the Dyson-Schwinger quark model
and various quark-gluon vertices}

\begin{CJK*}{GB}{gbsn} 

\author{H. Chen (³Â»¶)}
\email[]{huanchen@cug.edu.cn}
\author{J.-B. Wei (κ½ð±ê)}
\affiliation{
School of Mathematics and Physics, China University of Geosciences,
Lumo Road 388, 430074 Wuhan, China}

\author{
M. Baldo, G. F. Burgio, and H.-J. Schulze}
\affiliation{
INFN Sezione di Catania and Dipartimento di Fisica e Astronomia,
Universit\'a di Catania, Via Santa Sofia 64, 95123 Catania, Italy}


\begin{abstract}
We study cold dense quark matter and hybrid neutron stars with a
Dyson-Schwinger quark model and various choices of the quark-gluon vertex.
We obtain the equation of state of quark matter in beta equilibrium
and investigate the hadron-quark phase transition in combination with a
hadronic equation of state derived within the
Brueckner-Hartree-Fock many-body theory.
Comparing with the results for quark matter with the rainbow approximation,
the Ball-Chiu ansatz and the 1BC ansatz for the quark-gluon vertex
lead to a reduction of the effective interaction at finite chemical potential,
qualitatively similar to the effect of our gluon propagator.
We find that the phase transition and the equation of state of the quark
or mixed phase
and consequently the resulting hybrid star mass and radius
depend mainly on a global reduction of the effective interaction
due to effects of both the quark-gluon vertex and gluon propagator,
but are not sensitive to the vertex ansatz.
\end{abstract}


\maketitle

\end{CJK*}

\section{Introduction}

The possible appearance of quark matter (QM)
in the interior of massive neutron stars (NS)
is one of the main issues in the physics of compact stars \cite{gle}.
Recent observations confirm the existence of two NS of about
two solar masses \cite{heavy,heavy2}.
Based on a microscopic nucleonic equation of state (EOS),
one expects that in such heavy NS
the central particle density reaches values larger than $1/\text{fm}^3$,
where in fact quark degrees of freedom are expected to appear
at a macroscopic level.

The mass of a NS can be calculated by solving the Tolman-Oppenheimer-Volkoff
(TOV) equations with the relevant EOS as input,
which embodies the theoretical information of our theory on dense matter.
The hybrid EOS including both hadronic matter and QM is usually obtained
by combining EOSs of hadronic matter and QM within individual theories/models.
Unfortunately, while the microscopic theory of the nucleonic EOS has
reached a high degree of sophistication \cite{gle,bbb,akma,zhou,zhli,mmy},
the QM EOS is still poorly known at zero temperature and
at the high baryonic density appropriate for NS.

Continuing a set of investigations using different quark models
\cite{nsquark,njl,cdm,maru,roma},
in Ref.~\cite{Chen:2011}
we developed a Dyson-Schwinger model (DSM) for QM based on the
Dyson-Schwinger equations (DSE) of QCD
\cite{Roberts:1994dr,Alkofer:2000wg,Roberts:2000aa,Maris:2003vk,Roberts:2007jh}.
In that work we used the `rainbow' approximation, i.e.,
the bare quark-gluon vertex was employed.
In combination with a baryonic EOS developed within the Brueckner-Hartree-Fock
(BHF) many-body approach of nuclear matter,
we found that hybrid NS of two solar masses could be obtained
when hyperons were not included and the nuclear matter EOS was stiff enough.

Though the DSE are well based on QCD,
in practice one has to work within certain truncation schemes
of the infinite hierarchy of coupled equations.
In our truncation scheme for the DSE of the quark propagator,
we have thus to use an ansatz for the gluon propagator
and the quark-gluon vertex.
In this paper,
we investigate the effects of different choices for the quark-gluon vertex,
employing the so-called Ball-Chiu (BC) vertex \cite{ball-chiu}
or the 1BC vertex, i.e.,
the first term of BC vertex.
The comparison with the rainbow approximation will also be made.

The paper is organized as follows.
In section \ref{s:bhf} we briefly discuss the baryonic EOS in the BHF approach.
Section \ref{s:qm} concerns the DSM with the BC and 1BC ansatz.
In section \ref{s:res} we present the results regarding NS structure,
connecting the baryonic and QM EOS for beta-stable nuclear matter
with a phase transition under the Gibbs construction.
Section \ref{s:end} contains our conclusions.

\section{Formalism}

\subsection{Hadronic matter within Brueckner theory}
\label{s:bhf}

Our EOS of hadronic matter obtained within the BHF approach \cite{book}
has been amply discussed in previous publications \cite{Chen:2011}.
The basic input quantities of the calculation
are the nucleon-nucleon two-body potentials, namely
Argonne $V_{18}$ \cite{v18}, Bonn B \cite{bob}, or Nijmegen 93 \cite{n93},
supplemented with compatible three-body forces \cite{zhli,zuotbf,uix}.

This approach has also been extended with inclusion of hyperons
\cite{sch98,vi00,mmy},
which might appear in the core of a NS.
The hyperonic EOS in this theory is very soft,
which results in too low maximum masses of NS \cite{tom}
and often suppresses the appearance of quark matter.
In this work we do not discuss this aspect,
but choose a purely nucleonic EOS with the Bonn B potential,
supplemented by a compatible microscopic three-body force \cite{zhli}.
Since our main goal is to investigate the influence of our quark model,
and the inclusion of quark matter usually decreases the maximum mass of NS,
we select a hard EOS of hadronic matter,
which supports a large NS maximum mass.

The BHF calculations provide the energy density $\eps$ of the bulk system
as a function of the relevant partial densities $\rho_i$,
from which all other thermodynamical quantities can be obtained,
in particular chemical potentials and pressure,
\be
 \mu_i = {\partial \eps \over \partial \rho_i} \:,
\ee
\be
 p(\rob) = \rob^2 {d\over d\rob} {\eps\over\rob}
 = \rob {d\eps \over d\rob} - \eps
 = \rob \mu_B - \eps \:.
\ee
The parameterized energy density function can be found in Ref.~\cite{zhli}.

\subsection{Quark phase with the Dyson-Schwinger model}
\label{s:qm}

For the deconfined quark phase,
we adopt models based on the DSE of the quark propagator.
At finite chemical potential $\mu\equiv\mu_q=\mu_B/3$,
the quark propagator assumes a general form with rotational covariance,
\bea
 S(p;\mu)^{-1} &=&
 \left[ i{\bm \gamma}{\bm p} + i \gamma_4 (p_4+i\mu) + m_q \right]
 + \Sigma(p;\mu)
\\ &\equiv&
 i {\bm \gamma}{\bm p} \;A(p^2,p\cdot u) + B(p^2,p\cdot u)
\non\\&&
 + i \gamma_4(p_4+i\mu) \;C(p^2,p\cdot u) \:,
\label{sinvp}
\eea
where $m_q$ is the current quark mass,
$u=(\bm{0},i\mu)$,
and possibilities of other structures, e.g.,
color superconductivity \cite{alford,Yuan:2006yf,Nickel:2006vf},
are neglected.
The quark self-energy can be obtained from the gap equation,
\bea
 \Sigma(p;\mu) &=&
 \int\! \frac{d^4q}{(2\pi)^4} \,
 g^2(\mu) D_{\rho\sigma}(p-q;\mu)
\non\\&&\times
 \frac{\lambda^a}{2} \gamma_\rho S(q;\mu)
 \frac{\lambda^a}{2} \Gamma_\sigma(q,p;\mu) \:,
\label{gensigma}
\eea
where $\lambda^a$ are the Gell-Mann matrices,
$g(\mu)$ is the coupling strength,
$D_{\rho\sigma}(k;\mu)$ the dressed gluon propagator,
and $\Gamma_\sigma(q,p;\mu)$ the dressed quark-gluon vertex
at finite chemical potential.

This vertex is a complicated quantity,
which at zero chemical potential
can be decomposed into 12 linearly independent Lorentz covariants.
The DSE of the quark-gluon vertex depends on higher-order Green functions,
and so far little is known about it.
Commonly, one develops an ansatz constrained by all available reliable
information, e.g.,
the Slavnov-Taylor identities (STI),
effects of dynamical chiral symmetry breaking,
and some recent lattice results
\cite{Cloet:2014,Bashir:2012,Chang:2012,Rojas:2013,Aguilar:2014}.
In our previous work \cite{Chen:2011}
we employed the simple `rainbow' approximation,
$\Gamma_\sigma(q,p;\mu)=\gamma_\sigma$.

In this work, we will investigate another popular model,
the so-called BC vertex \cite{ball-chiu}.
The BC vertex satisfies the Ward-Takahashi identity (WTI) of QED,
which constrains the longitudinal part of the fermion-photon vertex,
and is free of kinetic singularity.
Furthermore, the scalar part of the BC vertex gives a direct representation
of the dynamical chiral symmetry breaking in the vertex.
Though the WTI is different from the STI of QCD,
this ansatz for the quark-gluon vertex is often used as a starting point
to constrain the vertex with STI of QCD \cite{Qin:2013}.
Phenomenologically, it also improves the results of
calculating hadron properties \cite{Roberts:2007jh,Chang:2009zb}.
It is widely used in comparison with the rainbow approximation
to identify robust features of a vertex ansatz.

The form of the BC vertex at finite chemical potential is developed
in \cite{Chen:2008zr},
\bea
 && i\Gamma^\text{BC}_\sigma(q,p;\mu) =
 i\Gamma^\text{1BC}_\sigma(q,p;\mu)
 + (\tilde q+\tilde p)_\sigma
 \Big[ \Delta_B(\tilde q,\tilde p;\mu)
\non\\&&
 + \frac{i}{2}\gamma^\perp \cdot (\tilde q+\tilde p)
 \Delta_A(\tilde q,\tilde p;\mu)
 + \frac{i}{2}\gamma^\| \cdot (\tilde q+\tilde p)
 \Delta_C(\tilde q,\tilde p;\mu) \Big]
\label{bcvtxmu}
\eea
with
\be
 i\Gamma^\text{1BC}_\sigma(q,p;\mu) =
 i\Sigma_A(q,p;\mu) \gamma^\perp_\sigma +
 i\Sigma_C(q,p;\mu) \gamma^\|_\sigma \:,
\label{1bcvtxmu}
\ee
where $\tilde q=q+u$, $\tilde p=p+u$,
and
$\gamma^\| = \hat u \gamma\cdot \hat u$,
$\gamma^\perp = \gamma - \gamma^\|$
with $\hat u^2=1$,
\bea
 \Sigma_F(q,p;\mu) &=& \frac{1}{2}
 \left[ F(\bm q^2,q_4;\mu) + F(\bm p^2,p_4;\mu) \right] \:,
\nonumber\\
 \Delta_F(\tilde q,\tilde p;\mu) &=&
 \frac{F(\bm q^2,q_4;\mu)-F(\bm p^2,p_4;\mu)}
 {\tilde q^2-\tilde p^2} \:,
\nonumber
\eea
with $F=A,B,C$ of Eq.~(\ref{sinvp}).
For comparison, we will also investigate the so-called 1BC vertex, i.e.,
the first term of Eq.~(\ref{bcvtxmu}) alone.

For the gluon propagator,
we still employ the Landau gauge form
with an infrared-dominant interaction
modified by the chemical potential \cite{Chen:2011,Jiang:2013}
\be
 g^2(\mu) D_{\rho \sigma}(k,\mu) =
 4\pi^2 d \frac{k^2}{\omega^6} e^{-\frac{k^2+\al\mu^2}{\omega^2}}
 \Big(\delta_{\rho\sigma}-\frac{k_\rho k_\sigma}{k^2}\Big) \:.
\label{gaussiangluonmu}
\ee
The various parameters can be obtained by fitting meson properties
and chiral condensate in vacuum \cite{Alkofer:2002bp,Chang:2009zb}
and we use
$\omega=0.5\;\text{GeV}$,
$d=1\;\text{GeV}^2$ (with the rainbow approximation),
$d=0.5\;\text{GeV}^2$ (with the BC vertex),
$d=0.75\;\text{GeV}^2$ (with the 1BC vertex),
$m_{u,d}=0$, and $m_s=0.115\;\text{GeV}$.
The phenomenological parameter $\al$ represents a reduction of the
effective interaction with increasing chemical potential.
This parameter cannot yet be fixed independently
and its value will be discussed in the following.

The EOS of cold QM is obtained following
Refs.~\cite{Chen:2011,Chen:2008zr,Klahn:2009mb}.
All the relevant thermodynamical quantities at zero temperature can be computed
from the quark propagator at finite chemical potential,
except a boundary value of the pressure,
which is represented by a phenomenological bag constant
$B_\text{DS}=90\;{\rm MeV\,fm^{-3}}$.
See Ref.~\cite{Chen:2011} for details.

\section{Results and discussion}
\label{s:res}

\subsection{EOS of dense matter in beta equilibrium}

In order to study the structure of NS,
we have to calculate the composition and the EOS
of cold, neutrino-free, charge-neutral, and beta-stable matter
characterized by two degrees of freedom $\mu_B$ and $\mu_e$,
the baryon and charge chemical potentials.
The corresponding equations are
\be
 \mu_i = b_i \mu_B - q_i \mu_e \:,\quad
 \sum_i \rho_i q_i = 0 \:,
\ee
$b_i$ and $q_i$ denoting baryon number and charge of
the particle species $i=n,p,e,\mu$ in the hadron phase
and $i=u,d,s,e,\mu$ in the quark phase, respectively.
At low density we assume a hadron phase and
will investigate a possible transition to quark matter at high density
under the Maxwell or Gibbs construction.

\begin{figure}[t]
\vspace{-19mm}
\includegraphics[scale=0.44]{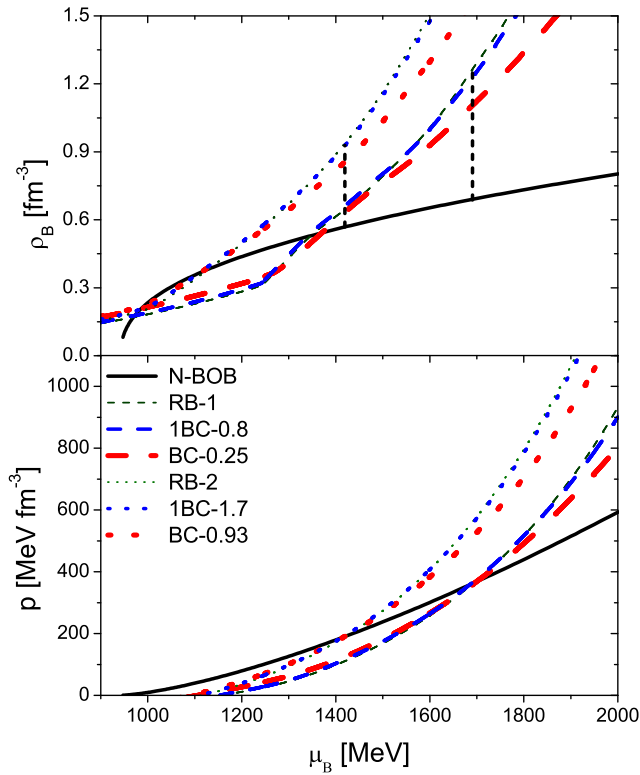}
\vspace{-14mm}
\caption{(Color online)
Baryon density (upper panel) and pressure (lower panel)
vs.~the baryon chemical potential of NS matter for different models.
The results for hadronic matter are shown with a black solid curve (N-BOB)
and for quark matter with curves of different thickness for the
rainbow approximation (RB-$\al$),
the 1BC vertex (1BC-$\al$), and the
BC vertex (BC-$\al$), respectively.
The vertical dotted lines indicate the positions of the phase transitions
under the Maxwell construction.}
\label{f:maxw}
\end{figure}

Fig.~\ref{f:maxw} shows
the numerical results for hadronic matter and quark matter in beta equilibrium
with different choices of the quark-gluon vertex and different parameters $\al$
given in Eq.~(\ref{gaussiangluonmu}).
One can directly read off the phase transition between hadron matter and quark
matter under the Maxwell construction,
which is presented as
crossing point of the baryon and quark pressure curves (lower panel) and the
projected vertical lines of the baryon density curves (upper panel).
Though the DSM EOS is generally stiffer than that with the MIT bag model
\cite{Chen:2011},
it strongly depends on the quark-gluon vertex and the parameter $\al$.
Within a given vertex ansatz,
the pressure and density increase as $\al$ increases.
For the same value of $\al$,
the pressure and density obtained with the 1BC or BC vertex
are larger than those with the rainbow approximation.
For the DSM with the rainbow approximation,
we repeat the results with $\al=1,2$ \cite{Chen:2011},
which were able to support hybrid stars with maximum masses
larger than two solar masses.

However, as stated above,
$\al$ is a phenomenological parameter that
currently cannot be fixed independently.
In the following, we will not compare the results between
different vertex \anss\ with the same value of $\al$,
but instead we constrain the parameter $\al$ with different vertex \anss\ by
requiring the same phase transition point under the Maxwell construction.
In other words,
we employ a global reduction rate of the interaction due to both the
gluon propagator and the quark-gluon vertex by fitting the same phase
transition point under the Maxwell construction.
This requires that $\al_\text{BC}<\al_\text{1BC}<\al_\text{RB}$.
In particular, we find
$\al=0.8(1.7)$ for the 1BC ansatz and $\al=0.25(0.93)$ for the BC ansatz
corresponding to $\al=1(2)$ for the rainbow approximation.
With such a choice of $\al$,
we obtain results with the rainbow approximation and
1BC ansatz that are almost indistinguishable,
while the pressure and densities
with the BC ansatz are only slightly lower as functions of chemical potential.

\begin{figure}[t]
\vspace{-7mm}
\includegraphics[scale=0.46]{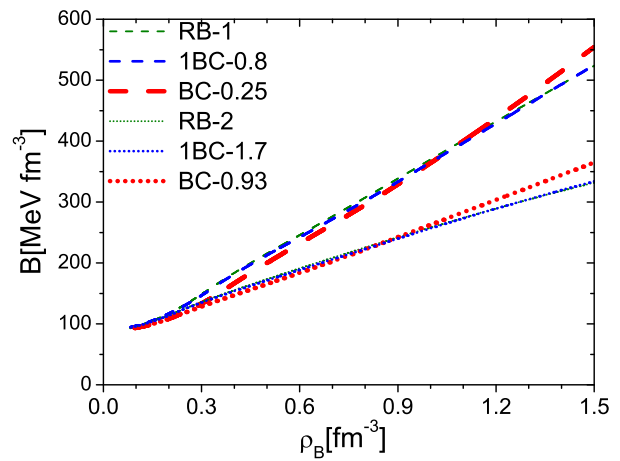}
\vspace{-8mm}
\caption{(Color online)
Effective bag constant of beta-stable quark matter
as a function of baryon number density for different quark models.}
\label{f:bag}
\end{figure}

A useful quantity to characterize an effective quark model is the
effective bag constant \cite{cdm,Chen:2011},
\be
 B(\rob) \equiv \eps(\rob) - \eps_\text{free}(\rob) \:,
\ee
defined as the difference of the energy densities of beta-stable quark matter
obtained for the DSM and free quark system, respectively.
It is shown in Fig.~\ref{f:bag}.
For all the three vertex \anss,
the effective bag constant is a monotonically increasing function of baryon
density,
presenting an important qualitative difference from the
generalized MIT bag model \cite{nsquark},
and are quantitatively much larger than the effective bag constants
from the color dielectric model \cite{cdm} and the Nambu-Jona-Lasinio model \cite{Buballa:04,Schertler:99}.
Once again, we find that also this quantity depends mainly on the
{\em rescaled} $\al$ parameter,
and not on the details of the vertex,
in which case
the rainbow approximation and the 1BC ansatz are almost indistinguishable,
while the results with the BC ansatz are only slightly higher at high densities.
With larger reduction rate of the
interaction the effective bag constant increase faster.

\begin{figure}[t]
\vspace{-16mm}
\includegraphics[scale=0.4]{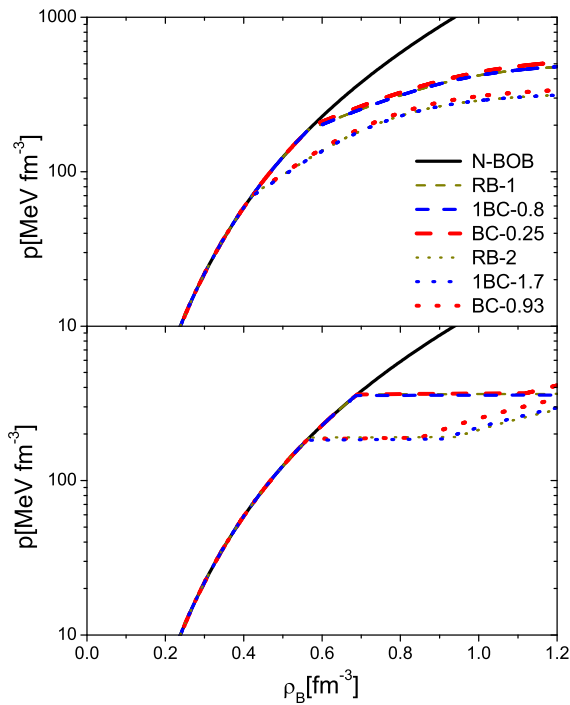}
\vspace{-11mm}
\caption{(Color online)
Pressure vs.~baryon density of NS matter
with the Gibbs (upper panel)
and Maxwell (lower panel)
phase transition construction for different quark models.}
\label{f:prho}
\end{figure}

We also investigate the phase transition under the
more sophisticated Gibbs construction \cite{gle,maru,glen},
as detailed in Eq.~(17-22) of Ref.~\cite{Chen:2011},
which comprises a mixed phase region
of oppositely charged hadron and quark matter domains,
while preserving the total charge neutrality.
The Gibbs construction represents the zero-surface-tension limit
of the calculations including finite-size effects \cite{maru,yasu,Chen:2013},
whereas the Maxwell construction corresponds to very large surface tension.
The resulting EOSs $p(\rob)$ are shown in the upper panel of Fig.~\ref{f:prho},
while the results under the Maxwell construction are shown in the lower panel
for comparison.
Again one observes that the density range and the EOS of the mixed phase
depend mainly on the global reduction rate,
and are then almost indistinguishable with the three vertex \anss.

\begin{figure}[t]
\vspace{-2mm}
\includegraphics[bb=30 90 513 610,scale=0.5]{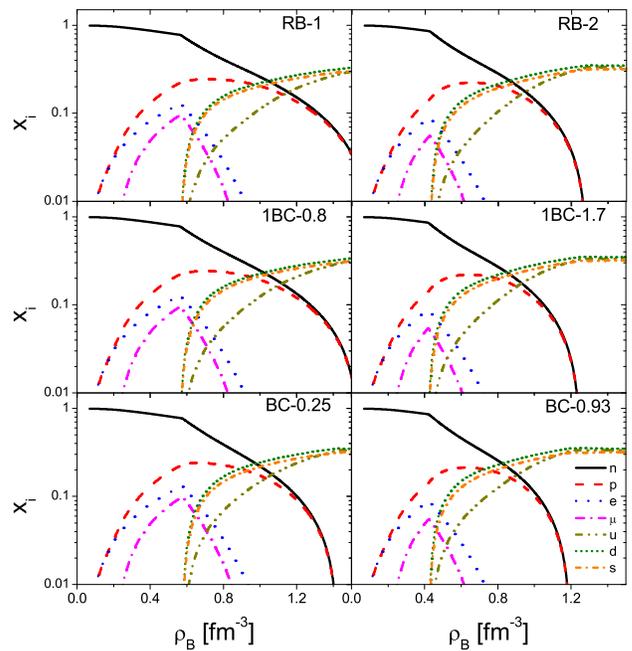}
\vspace{-4mm}
\caption{(Color online)
Fractions of all particle species as functions of baryon number density
with the Gibbs phase transition construction
for different quark models.}
\label{f:frac}
\end{figure} 

Fig.~\ref{f:frac} shows more detailed information of the mixed phase,
namely the fractions
$x_i \equiv {\bar\rho_i}/{\rob}$
of all species of particles
as functions of baryon number density,
where in the mixed phase
with quark volume fraction $\chi$
the average partial densities are given by
$\bar\rho_i = \chi {\rho_i}$ for quarks and
$\bar\rho_i = (1-\chi){\rho_i}$ for hadrons.
Also these quantities are almost the same for different vertex \anss,
once the properly scaled parameter $\al$ is chosen.

\subsection{Neutron star structure}

As usual, we assume that a NS is a spherically symmetric distribution of
mass in hydrostatic equilibrium and
obtain the stellar radius $R$ and the gravitational mass $M$
by the standard process of solving the TOV equations \cite{shapiro}.
We have used as input the EOSs with the Gibbs construction discussed above
and shown in the upper panel of Fig.~\ref{f:prho}.
The hybrid NS with phase transitions under the Maxwell construction
are all unstable and the corresponding results are not shown here.
For the description of the NS crust, we have joined the hadronic EOS with the
ones by Negele and Vautherin \cite{nv} in the medium-density regime,
and the ones by Feynman-Metropolis-Teller \cite{fey} and
Baym-Pethick-Sutherland \cite{bps} for the outer crust.

\begin{figure}[t]
\vspace{-6mm}
\includegraphics[scale=0.48]{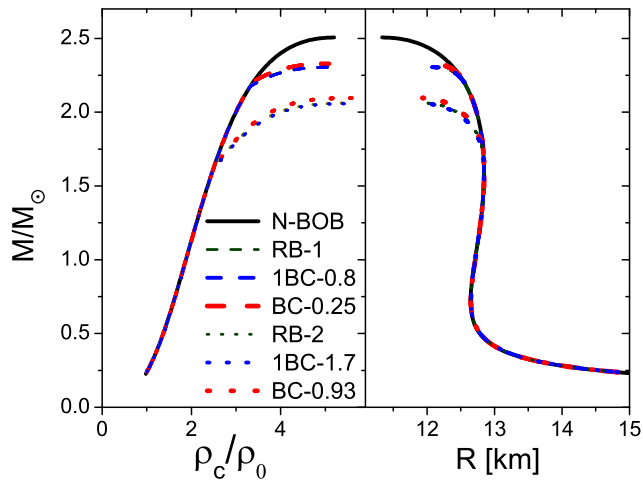}
\vspace{-7mm}
\caption{(Color online)
Gravitational NS mass
($M_\odot=2\times10^{33}\text{g}$)
vs.~radius (right panel)
and normalized central baryon density
(left panel; $\rho_0=0.17\;\text{fm}^{-3}$)
for different EOSs.}
\label{f:mr}
\end{figure}

The results are plotted in Fig.~\ref{f:mr},
where we display the gravitational mass $M$
as a function of the central baryon number density $\rho_{c}$
and the radius $R$.
For all the three vertex \anss\ and the chosen values of $\al$,
we obtain hybrid stars with the maximum mass lower than the pure nucleonic NS,
but higher than two solar masses.
From Figs.~\ref{f:frac} and \ref{f:mr} one deduces
that even in the most massive hybrid stars no pure quark phase exists,
but only a mixed phase of nuclear matter and QM in the inner core.
Consistent with the results for the EOS,
also the hybrid NS mass-radius relation depends mainly on the scaling parameter
$\al$ and does then not allow to distinguish between the different choices
of the quark-gluon vertex.
To get a better understanding of the quark-gluon vertex in the DSM,
one therefore needs refined and independent information
about the parameter $\al$,
i.e., the in-medium modification of the gluon propagator.

\section{Conclusions}
\label{s:end}

We have further developed our DSM for QM under the BC and 1BC \anss\ for the
quark-gluon vertex in comparison with the previous rainbow approximation.
We investigated the EOS of beta-stable quark matter in NS
and the phase transition
between hadronic matter and QM in combination with a microscopic hadronic
EOS obtained within the BHF formalism.

Due to the uncertainty of the in-medium gluon propagator in the DSM,
which we currently approximate by a single parameter $\al$,
we obtain a global modification of the interaction caused by both
the quark-gluon vertex and the gluon propagator.
The freedom of the parameter $\al$ allows then
to fit the same phase transition point under the Maxwell construction
for the different choices of the vertex.
Consequently, we find that all observables, i.e.,
EOS, particle fractions, and stellar structure,
mainly depend on the global reduction rate of the interaction,
but are not sensitive to the particular ansatz for the quark-gluon vertex.
With a proper choice of $\al$ and for all three vertex \anss,
the EOS from the DSM can support a two-solar-mass hybrid star,
provided that the appearance of hyperons is excluded.

These results focus the attention to the necessity of obtaining
model-independent information about the quark-gluon vertex
and gluon propagator at finite baryon density.
More investigation on the quark-gluon vertex
and gluon propagator from their own DSEs will be done in the future.

\section*{Acknowledgments}

We acknowledge financial support from NSFC (11305144),
Central Universities (GUGL 130605,140609) in China.
Partial support comes from ``NewCompStar," COST Action MP1304.



\begin{thebibliography}{99}

\bibitem{gle}
 N. K. Glendenning,
 {\em Compact Stars, Nuclear Physics, Particle Physics, and General Relativity},
 2nd ed., 2000, Springer-Verlag, New York.

\bibitem{heavy}
 P. B. Demorest, T. Pennucci, S. M. Ransom, M. S. E. Roberts,
 and J. W. T. Hessels,
 Nature {\bf 467}, 1081 (2010).

\bibitem{heavy2}
 J. A. Antoniadis et al.,
 Science {\bf 340}, 6131 (2013).

\bibitem{bbb}
 M. Baldo, I. Bombaci, and G. F. Burgio,
 Astron. Astrophys. {\bf 328}, 274 (1997).

\bibitem{akma}
 A. Akmal, V. R. Pandharipande, and D. G. Ravenhall,
 Phys. Rev. {\bf C58}, 1804 (1998).

\bibitem{zhou}
 X. R. Zhou, G. F. Burgio, U. Lombardo, H.-J. Schulze, and W. Zuo,
 Phys. Rev. {\bf C69}, 018801 (2004).

\bibitem{zhli}
 Z. H. Li and H.-J. Schulze,
 Phys. Rev. {\bf C78}, 028801 (2008).

\bibitem{mmy}
 H.-J. Schulze, A. Polls, A. Ramos, and I. Vidana,
 Phys. Rev. {\bf C73}, 058801 (2006).

\bibitem{nsquark}
 G. F. Burgio, M. Baldo, P. K. Sahu, and H.-J. Schulze,
 Phys. Rev. {\bf C66}, 025802 (2002).

\bibitem{njl}
 M. Baldo, M. Buballa, G. F. Burgio, F. Neumann, M. Oertel, and H.-J. Schulze,
 Phys. Lett. {\bf B562}, 153 (2003);
 M. Baldo, G. F. Burgio, P. Castorina, S. Plumari, and D. Zappal\`a,
 Phys. Rev. {\bf C75}, 035804 (2007).

\bibitem{cdm}
 C. Maieron, M. Baldo, G. F. Burgio, and H.-J. Schulze,
 Phys. Rev. {\bf D70}, 043010 (2004).

\bibitem{maru}
 T. Maruyama, S. Chiba, H.-J. Schulze, and T. Tatsumi,
 Phys. Rev. {\bf D76}, 123015 (2007).

\bibitem{roma}
 A. Kurkela, P. Romatschke, and A. Vuorinen,
 Phys. Rev. {\bf D81}, 105021 (2010).

\bibitem{Chen:2011}
 H. Chen, M. Baldo, G. F. Burgio, and H.-J. Schulze,
 Phys. Rev. {\bf D84}, 105023 (2011);
 {\bf D86}, 045006 (2012).

\bibitem{Roberts:1994dr}
 C. D. Roberts and A. G. Williams,
 Prog. Part. Nucl. Phys. {\bf 33}, 477 (1994).

\bibitem{Alkofer:2000wg}
 R. Alkofer and L. von Smekal,
 Phys. Rep. {\bf 353}, 281 (2001).

\bibitem{Roberts:2000aa}
 C. D. Roberts and S. M. Schmidt,
 Prog. Part. Nucl. Phys. {\bf 45}, S1 (2000).

\bibitem{Maris:2003vk}
 P. Maris and C. D. Roberts,
 Int. J. Mod. Phys. {\bf E12}, 297 (2003).

\bibitem{Roberts:2007jh}
 C. D. Roberts, M. S. Bhagwat, A. H\"oll, and S. V. Wright,
 Eur. Phys. J. Special Topics {\bf 140}, 53 (2007).

\bibitem{ball-chiu}
 J. S. Ball and T. W. Chiu,
 Phys. Rev. {\bf D22}, 2542,2550 (1980); {\bf D23}, 3085 (1981).

\bibitem{book}
 M. Baldo,
 {\em Nuclear Methods and the Nuclear Equation of State},
 International Review of Nuclear Physics, Vol. 8
 (World Scientific, Singapore, 1999).

\bibitem{v18}
 R. B. Wiringa, V. G. J. Stoks, and R. Schiavilla,
 Phys. Rev. {\bf C51}, 38 (1995).

\bibitem{bob}
 R. Machleidt, K. Holinde, and Ch. Elster,
 Phys. Rep. {\bf 149}, 1 (1987);
 R. Machleidt,
 Adv. Nucl. Phys. {\bf 19}, 189 (1989).

\bibitem{n93}
 M. M. Nagels, T. A. Rijken, and J. J. de Swart,
 Phys. Rev. {\bf D17}, 768 (1978);
 V. G. J. Stoks, R. A. M. Klomp, C. P. F. Terheggen, and J. J. de Swart,
 Phys. Rev. {\bf C49}, 2950 (1994).

\bibitem{zuotbf}
 A. Lejeune, P. Grang\a'{e}, M. Martzolff, and J. Cugnon,
 Nucl. Phys. {\bf A453}, 189 (1986);
 W. Zuo, A. Lejeune, U. Lombardo, and J.-F. Mathiot,
 Nucl. Phys. {\bf A706}, 418 (2002);
 Z. H. Li, U. Lombardo, H.-J. Schulze, and W. Zuo,
 Phys. Rev. {\bf C77}, 034316 (2008).

\bibitem{uix}
 J. Carlson, V. R. Pandharipande, and R. B. Wiringa,
 Nucl. Phys. {\bf A401}, 59 (1983);
 R. Schiavilla, V. R. Pandharipande, and R. B. Wiringa,
 Nucl. Phys. {\bf A449}, 219 (1986);
 B. S. Pudliner, V. R. Pandharipande, J. Carlson, S. C. Pieper, and R. B. Wiringa,
 Phys. Rev. {\bf C56}, 1720 (1997).

\bibitem{sch98}
 H.-J. Schulze, M. Baldo, U. Lombardo, J. Cugnon, and A. Lejeune,
 Phys. Rev. {\bf C57}, 704 (1998);
 M. Baldo, G. F. Burgio, and H.-J. Schulze,
 Phys. Rev. {\bf C58}, 3688 (1998);
 Phys. Rev. {\bf C61}, 055801 (2000).

\bibitem{vi00}
 I. Vida\~na, A. Polls, A. Ramos, M. Hjorth-Jensen, and V. G. J. Stoks,
 Phys. Rev. {\bf C61}, 025802 (2000).

\bibitem{tom}
 H.-J. Schulze and T. Rijken,
 Phys. Rev. {\bf C84}, 035801 (2011).

\bibitem{alford}
 M. Alford and S. Reddy,
 Phys. Rev. {\bf D67}, 074024 (2003).

\bibitem{Yuan:2006yf}
 W. Yuan, H. Chen, and Y.-X. Liu,
 Phys. Lett. {\bf B637}, 69 (2006).

\bibitem{Nickel:2006vf}
 D. Nickel, J. Wambach, and R. Alkofer,
 Phys. Rev. {\bf D73}, 114028 (2006);
 D. Nickel, R. Alkofer, and J. Wambach,
 Phys. Rev. {\bf D74}, 114015 (2006).

\bibitem{Cloet:2014}
 I. C. Clo\"et and C. D. Roberts,
 Prog. Part. Nucl. Phys. {\bf 77}, 1 (2014).

\bibitem{Bashir:2012}
 A. Bashir, R. Bermudez, L. Chang, and C. D. Roberts,
 Phys. Rev. {\bf C85}, 045205 (2012).

\bibitem{Chang:2012}
 L. Chang and C. D. Roberts,
 Phys. Rev. {\bf C85}, 052201(R) (2012).

\bibitem{Rojas:2013}
 E. Rojas, J. de Melo, B. El-Bennich, O. Oliveira, and T. Frederico,
 JHEP {\bf 1310}, 193 (2013).

\bibitem{Aguilar:2014}
 A. C. Aguilar, D. Binosi, D. Iba\"nez, and J. Papavassiliou,
 Phys. Rev. {\bf D90}, 065027 (2014).

\bibitem{Qin:2013}
 S.-X. Qin, L. Chang, Y.-X. Liu, C. D. Roberts, and S. M. Schmidt,
 Phys. Lett. {\bf B722}, 384 (2013).

\bibitem{Chang:2009zb}
 L. Chang and C. D. Roberts,
 Phys. Rev. Lett. {\bf 103}, 081601 (2009).

\bibitem{Chen:2008zr}
 H. Chen, W. Yuan, L. Chang, Y. X. Liu, T. Kl\"ahn, and C. D. Roberts,
 Phys. Rev. {\bf D78}, 116015 (2008).

\bibitem{Jiang:2013}
 Y. Jiang, H. Chen, W.-M Sun, and H.-S. Zong,
 JHEP {\bf 04}, 014 (2013).

\bibitem{Alkofer:2002bp}
 R. Alkofer, P. Watson, and H. Weigel,
 Phys. Rev. {\bf D65}, 094026 (2002).

\bibitem{Klahn:2009mb}
 T. Kl\"ahn, C. D. Roberts, L. Chang, H. Chen, and Y. X. Liu,
 Phys. Rev. {\bf C82}, 035801 (2010).

\bibitem{Buballa:04}
 M. Buballa, Phys. Rept. {\bf 407}, 205 (2005).

\bibitem{Schertler:99}
K. Schertler, S. Leupold, and J. Schaffner-Bielich, Phys.
Rev. {\bf C60}, 025801 (1999).

\bibitem{glen}
 N. K. Glendenning,
 Phys. Rev. {\bf D46}, 1274 (1992).

\bibitem{yasu}
 N. Yasutake, T. Maruyama, and T. Tatsumi,
 Phys. Rev. {\bf D86}, 101302 (2012).

\bibitem{Chen:2013}
 H. Chen, G. F. Burgio, H.-J. Schulze, and N. Yasutake,
 Astron. Astrophys. {\bf 551} A13 (2013).

\bibitem{shapiro}
 S. L. Shapiro and S. A. Teukolsky,
 {\em Black Holes, White Dwarfs, and Neutron Stars}
 (John Wiley \& Sons, New York, 1983).

\bibitem{nv}
 J. W. Negele and D. Vautherin,
 Nucl. Phys. {\bf A207}, 298 (1973).

\bibitem{fey}
 R. P. Feynman, N. Metropolis, and E. Teller,
 Phys. Rev. {\bf 75}, 1561 (1949).

\bibitem{bps}
 G. Baym, C. Pethick, and D. Sutherland,
 Astrophys. J. {\bf 170}, 299 (1971).

\end{thebibliography}
\end{document}